\documentclass{iau}
\usepackage{graphicx}

\title[Magnetic fields across the PMS of the H-R diagram] 
{Can we predict the magnetic properties of PMS stars from their H-R diagram location?}

\author[S. G. Gregory et al.]   
{S. G. Gregory$^1$, J.-F. Donati$^2$, J. Morin$^3$, G. A. J. Hussain$^4$, \\ 
N. J. Mayne$^5$, L. A. Hillenbrand$^6$ 
 \and M. Jardine$^1$}

\affiliation{$^1$School of Physics \& Astronomy, University of St Andrews, St Andrews,
KY16 9SS, U.K. \\ email: {\tt sg64@st-andrews.ac.uk} \\[\affilskip]
$^2$UPS-Toulouse/CNRS-INSU, IRAP UMR 5277, Toulouse, FÐ31400 France \\[\affilskip]
$^3$Inst. f{\"u}r Astrophysik, Univ. G{\"o}ttingen, Friedrich-Hund-Platz 1, D-37077 G{\"o}ttingen, Germany \\[\affilskip]
$^4$ESO, Karl-Schwarzschild-Str. 2, D-85748 Garching, Germany \\[\affilskip]
$^5$School of Physics, University of Exeter, Exeter EX4 4QL, U.K. \\[\affilskip]
$^6$California Institute of Technology, MC 249-17, Pasadena, CA 91125, U.S.A. \\[\affilskip]
}

\pubyear{2013}
\volume{302}  
\pagerange{xxx--xxx}
\jname{Magnetic fields throughout stellar evolution}
\editors{M. Jardine, P. Petit \& H. Spruit, eds.}
\begin{document}

\maketitle

\begin{abstract}
Spectropolarimetric observations combined with tomographic imaging techniques have revealed
that all pre-main sequence (PMS) stars host multipolar magnetic fields, ranging from strong and globally axisymmetric with $\gtrsim$kilo-Gauss dipole components,
to complex and non-axisymmetric with weak dipole components ($\lesssim$0.1\,{\rm kG}).  Many host dominantly octupolar large-scale fields.
We argue that the large-scale magnetic properties of a PMS star are related to its location in the Hertzsprung-Russell diagram. This conference paper is a synopsis of \cite[Gregory et al. (2012)]{gre12}, updated to include the latest results from magnetic mapping studies of PMS stars. 
\keywords{stars: evolution, stars: interiors, stars: magnetic field, stars: pre-main sequence}
\end{abstract}

\firstsection 
 \vspace*{-0.1 cm}              
\section{Introduction}
Since the 1990s it has been known that pre-main sequence (PMS) stars are capable of generating magnetic fields of a few kilo-Gauss in strength (e.g. \cite[Basri et al. 1992]{bas92}; \cite[Johns-Krull et al. 1999]{joh99}; \cite[Johns-Krull 2007]{joh07}; \cite[Yang \& Johns-Krull 2011]{yan11}). Fields of this magnitude, provided that they are sufficiently globally ordered, are easily strong enough to truncate circumstellar disks during the classical T Tauri star phase (e.g. \cite[K{\"o}nigl 1991]{kon91}).
        
Zeeman-Doppler imaging, combined with the techniques of least squares deconvolution (LSD; \cite[Donati et al. 1997]{don97}) and tomographic imaging, allows magnetic maps to be derived from circularly polarised spectra.  
For accreting PMS stars, the maps are constructed by simultaneously considering the rotationally modulated polarisation signature in both the LSD-averaged photospheric line and in the accretion-related emission lines (\cite[Donati et al. 2010b]{don10b}).  The maps themselves can then be decomposed into the various $\ell$ and $m$-number spherical harmonic modes (where $\ell=1,2,3\ldots$ are the dipole, quadrupole, octupole$\ldots$ field components). 

The first magnetic maps of an accreting PMS star were published by \cite[Donati et al. (2007)]{don07}, and have since been obtained for a small sample of stars - see the tables in \cite[Gregory et al. (2012)]{gre12} for a list which has recently been expanded to include DN~Tau (\cite[Donati et al. 2013]{don13}). Most PMS star magnetic maps have been derived during the MaPP (Magnetic Protostars \& Planets) project (PI: J.-F. Donati).  The data acquisition phase of MaPP (2008-12) with ESPaDOnS at the Canada-France-Hawai'i Telescope (CFHT) and NARVAL at the T{\'e}lescope Bernard Lyot, is now complete.  A highlight of MaPP is the discovery of a clear link between the internal structure of a star and
its external, large-scale, magnetic field topology. 

\begin{figure}[t]
 \vspace*{-0.2 cm}
\begin{center}
 \includegraphics[width=0.385\textwidth]{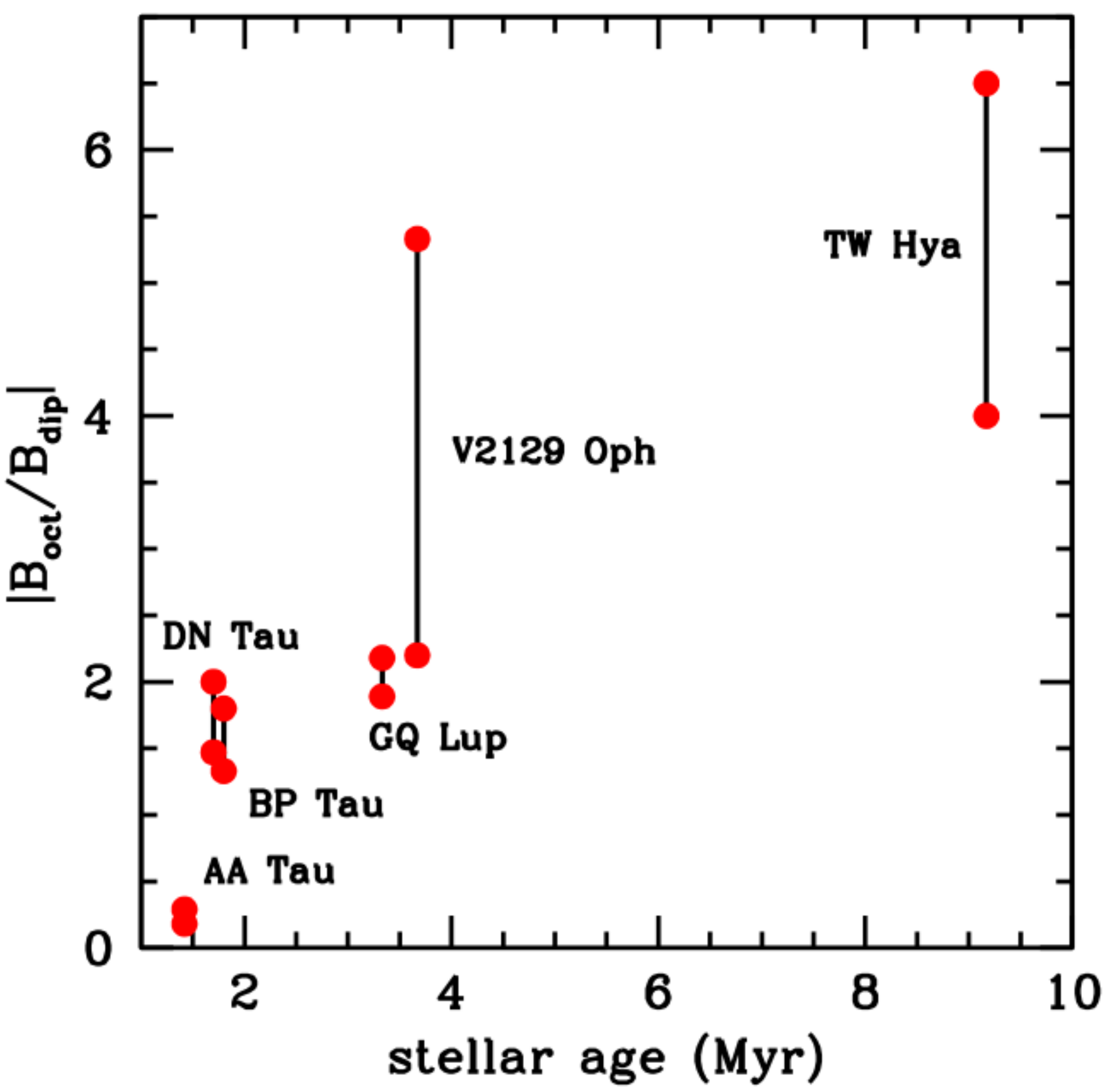}
 \includegraphics[width=0.381\textwidth]{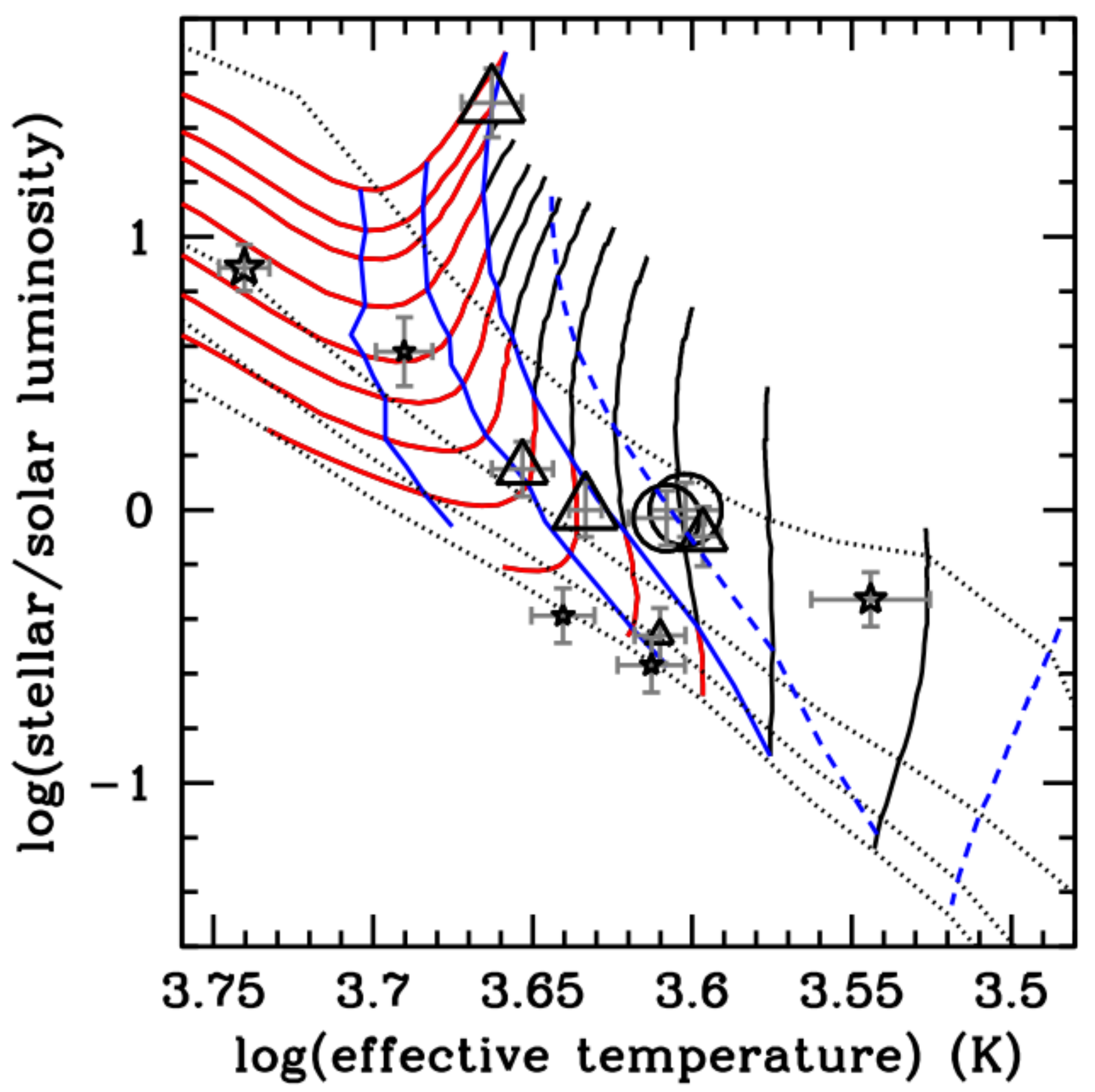}
 \vspace*{-0.2 cm}
 \caption{[left] Almost all fully convective plus partially convective PMS stars with small radiative cores with published magnetic maps have dipole plus octupole fields. The magnitude of the polar strength of the dipole to the octupole increases with age. All stars have been observed at least twice, at epochs one-to-a-few years apart. [right] A H-R diagram showing stars with published magnetic maps (updated from \cite[Gregory et al. 2012]{gre12}).  Mass tracks (solid black/red during the fully/partially convective phase of evolution for $M_\ast/{\rm M}_\odot=0.3-1.9$ in steps of $0.2$, then $2.2, 2.5, 2.7$ \& $3.0$ from right to left) and isochrones (dotted lines; age $=1, 5, 10$ \& $15\,{\rm Myr}$ from upper right to lower left) from \cite[Siess et al (2000)]{sie00} are shown.  Solid blue lines are the fully convective limit (right), and the loci of stars with radiative core masses of $M_{\rm core}/M_\ast=0.4$ \& $0.8$ (middle \& left).  Circles are fully convective stars with axisymmetric large-scale fields with strong ($\sim$kG) dipole components.  Triangles are stars with small radiative cores ($M_{\rm core}/M_\ast<0.4$; with the exception of DN~Tau, see section 2) and large-scale fields that are mostly axisymmetric and (typically) dominantly octupolar.  Asterisks are stars with non-axisymmetric large-scale magnetic fields with weak dipole components ($\lesssim$0.1$\,{\rm kG}$). Dashed blue lines are discussed in section 3.}
   \label{dipoctage}
\end{center}
\end{figure}

 \vspace*{-1.0 cm}
\section{PMS star magnetic topology \& the link with stellar structure}
About half of the accreting PMS stars with published magnetic maps have large-scale fields that are well described by slightly tilted dipole and octupole components (other field modes are present too but in almost all cases are less significant; \cite[Gregory \& Donati 2011]{gre11}).  
AA~Tau, BP~Tau, DN~Tau, GQ~Lup, TW~Hya
\& V2129~Oph have this sort of magnetic topology (\cite[Donati et al. 2007, 2008b, 2010b, 2011a,c, 2012, 2013]{don07,don08,don10b,don11a,don11c,don12,don13}), with the first (last) three listed having fully convective interiors (small radiative cores, $M_{\rm core}/M_\ast$$\lesssim$0.4; see \cite[Gregory et al. 2012]{gre12}). Their large-scale fields are dominantly axisymmetric and it appears that the magnitude of the ratio of the polar strength of the octupole to the dipole component $|B_{\rm oct}/B_{\rm dip}|$ increases with age (Fig.\,\ref{dipoctage}).  Fully convective stars are capable of generating strong kG dipole components, while those with small radiative cores have dominantly octupolar magnetic fields with dipole components that vary from a few times 0.1\,kG to of order $\sim$kG. The exception to this is the fully convective DN~Tau \cite[(Donati et al. 2013)]{don13}, although its $|B_{\rm oct}/B_{\rm dip}|$ ratio, and field polarity distribution across the stellar surface, is similar to that of the other fully convective stars (see below regarding dynamo bistability). Intriguingly, although the large-scale fields of PMS stars are evolving between observing epochs (e.g. \cite[Donati et al. 2011a]{don11a}), their general magnetic topology features remain unchanged across year-long timescales e.g. if a star has a dominantly octupolar magnetic field at one epoch,      
the field still has this configuration at the next epoch.

The other half of the sample have complex non-axisymmetric large-scale magnetic 
fields with weak dipole components ($\lesssim$0.1$\,{\rm kG}$).  CR~Cha, CV~Cha, V2247~Oph, V4046 Sgr A \& V4046~Sgr B have such large-scale magnetic fields (\cite[Hussain et al. 2009]{hus09}; \cite[Donati et al. 2010a, 2011b]{don10a,don11b}). All apart from V2247~Oph have large radiative cores ($M_{\rm core}/M_\ast\gtrsim0.4$).     

A H-R diagram showing stars with published magnetic maps is 
shown in Fig.\,\ref{dipoctage}.  The symbol type is related to the
large-scale field topology of the star, as described in the figure caption.  
\cite[Donati et al. (2011c)]{don11c} and \cite[Gregory et al. (2012)]{gre12} argued that there appears to be evidence for a magnetic evolutionary scenario. {\it Fully convective stars, which lie close to the fully convective limit, host simple axisymmetric large-scale fields and can have strong ($\sim$kG) dipole components.  The octupole component becomes more and more dominant once a radiative core develops, with the large-scale magnetic field eventually becoming complex and non-axisymmetric with a weak dipole 
($\sim$0.1$\,$kG) once the core has grown to occupy a sufficient proportion of the stellar interior (empirically once $M_{\rm core}/M_\ast\gtrsim0.4$).}    


\vspace*{-0.5 cm}
\section{The magnetic Hertzsprung-Russell diagram \& discussion}
There is a limited sample of PMS stars with published magnetic maps.  However, main sequence (MS) M-dwarfs follow similar magnetic topology trends (\cite[Donati et al. 2008a; Morin et al. 2008]{mor08,don08b}), with a weakening dipole component as we consider stars that have more and more radiative interiors. Their 
large-scale magnetic fields remain dominantly axisymmetric unless the radiative core mass exceeds $\sim$40\% of the total stellar mass 
\cite[(Gregory et al. 2012)]{gre12}; similar to what is found for the PMS star sample. 

\begin{figure}[t]
 \vspace*{-0.3 cm}
\begin{center}
 \includegraphics[width=0.78\textwidth]{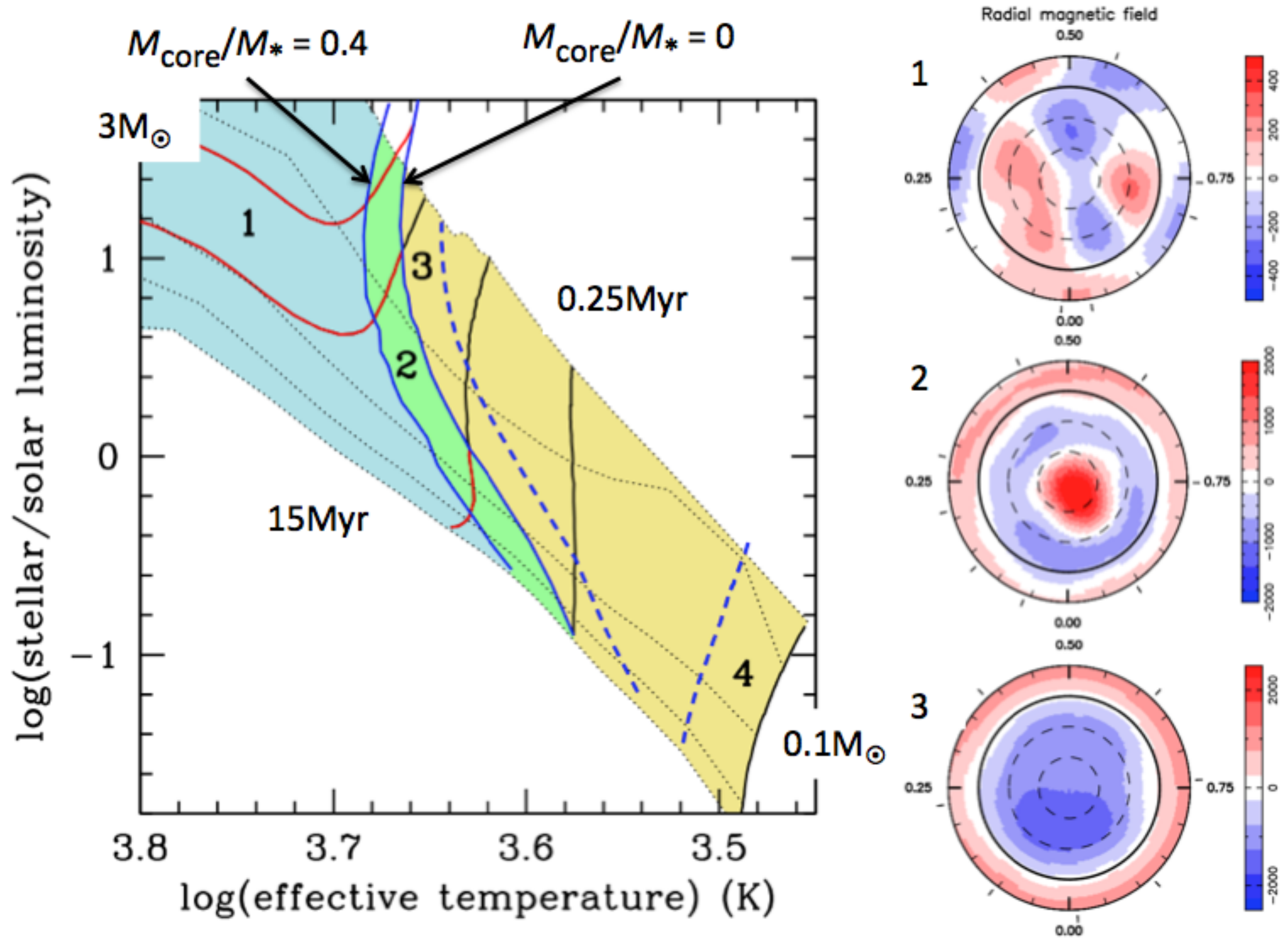} 
 \vspace*{-0.2 cm}
 \caption{A magnetic H-R diagram (\cite[Gregory et al. 2012]{gre12}).  The PMS is shown with mass tracks coloured as in Fig.\,\ref{dipoctage} (0.1, 0.5, 1, 2 \& 3 ${\rm M}_\odot$) and isochrones (0.25, 1, 5, 10 \& 15 $\,{\rm Myr}$) from \cite[Siess et al. (2000)]{sie00}.  Stars in the different numbered regions have differing internal structures and are observed to have different large-scale magnetic field topologies. Solid blue lines connect the loci of effective temperatures and luminosities where stars have the same internal structure, defined as the same ratio of radiative core mass to stellar mass. Stars in region 1 are largely radiative and host complex non-axisymmetric large-scale magnetic fields.  Stars in region 2 have small radiative cores ($0\le M_{\rm core}/M_\ast\le0.4$) and axisymmetric magnetic fields that are (typically) dominantly octupolar. Stars in region 3 are fully convective and host axisymmetric fields with kilo-Gauss dipole components. A fourth magnetic topology region exists at the lowest masses (see text) with the dashed blue lines indicating upper/lower limits to the boundary between regions 3 \& 4. The magnetic maps (right column) from top-to-bottom are V4046 Sgr B, V2129 Oph \& AA Tau (\cite[Donati et al. 2011b,a, 2010b]{don11b,don11a,don10b}). They are shown in flattened polar projection. Blue/red denotes negative/positive field with fluxes labelled in Gauss. Dashed lines are lines of constant latitude, separated by 30$^\circ$, with the bold circle the stellar equator. From top-to-bottom the stars are largely radiative with a complex multipolar magnetic field; partially convective with a small core and a dominantly octupolar field; and fully convective with a simple dipole magnetic field.}
   \label{maghrd}
\end{center}
\vspace*{-0mm}
\end{figure}

The similarities between the MS and PMS star samples prompted \cite[Gregory et al. (2012)]{gre12} to ask if we can predict the large-scale magnetic properties of a PMS star solely from its H-R diagram location. For example, will its large-scale field be dominantly axisymmetric or non-axisymmetric? Will it be dominantly octupolar, or more complex? Will the dipole component be of order $\sim$kG or $\sim$0.1$\,$kG? There appears to be at least three (defined in the caption of Fig.\,\ref{maghrd}) magnetic topology regimes across the PMS of the H-R diagram.      

A fourth, bistable dynamo, regime may exist amongst the lowest mass fully convective PMS stars, similar to that discovered for late M-dwarfs (\cite[Morin et al. 2011]{mor11}).  Amongst the fully convective PMS stars, DN~Tau hosts a large-scale field that is dominantly octupolar (see section 2) while the large-scale field of V2247~Oph is more akin to those of substantially radiative PMS stars (\cite[Donati et al. 2010a]{don10a}).  Thus, \cite[Gregory et al. (2012)]{gre12} speculated that a bistable dynamo regime exists for the lowest mass fully convective PMS stars, where stars with a variety of large-scale magnetic field topologies will be discovered. This is labelled as region 4 in Fig.\,\ref{maghrd} - the dashed blue lines show possible upper/lower limits below which bistable behaviour may be found. For MS stars dynamo bistability is found at $\lesssim$0.2$\,{\rm M}_\odot$, which sets the lower limit.  The upper limit is set by noting that $0.2\,{\rm M}_\odot$ is $\sim$60\% of the MS fully convective limit of $0.35\,{\rm M}_\odot$.  As the fully convective limit is a function of age (see \cite[Gregory et al. 2012]{gre12}), then the upper limit corresponds to a mass of 60\% of the fully convective limit at a given age. DN~Tau falls between the limits in the H-R diagram, and suggests that there may be a smooth transition from simple magnetic fields amongst the more massive fully convective stars, to the bistable dynamo regime of lower mass stars.   

A more complete magnetic mapping survey (of fully convective PMS stars in particular) is required to confirm or refute the arguments herein, a task ideally suited for SPIRou, the under construction nIR spectropolarimeter for CFHT. The clearest magnetic topology trend is that PMS stars with shallow convective zone depths have far more complex large-scale magnetic fields with dipole components up to an order of magnitude below those found for those with deep convective zones.  

{\it Acknowledgements}:
SGG acknowledges support from the Science \& Technology Facilities Council (STFC) via an Ernest Rutherford Fellowship [ST/J003255/1].

\end{document}